\documentclass[a4paper]{elsart}
\usepackage{natbib}
\usepackage{epsfig}

\begin{document}

\begin{frontmatter}

\title{A compact, multi-pixel parametric light source}
\author[Barna]{Stefano Minardi}
\author[Vilnius]{Arunas Varanavi\v{c}us} 
\author[Vilnius]{Algis Piskarskas}
\author[Como]{Paolo Di Trapani}

\address[Barna]{Instituto de Cinecias Fotonicas, c/Jordi Girona, 29 - NEXUS II, E-08034 Barcelona, Spain}
\address[Vilnius]{Department of Quantum Electronics, Vilnius University,
Sauletekio 9, LT-2040 Vilnius, Lithuania}
\address[Como]{Istituto Nazionale Fisica della Materia and Department of
Sciences, University of Insubria, Via Valleggio 11, I-22100 Como, Italy}

\begin{abstract}
The features of a compact, single pass, multi-pixel optical parametric generator
are discussed. Several hundreds of independent high spatial-quality tunable ultrashort pulses were produced by pumping a bulk lithium triborate crystal with an array of tightly focussed intense beams. The array of beams was produced by shining a microlenses array with a large pump beam. Overall conversion efficiency to 
signal and idler up to 30\% of the pump beam has been reported. Shot-to-shot 
energy fluctuation down to 3\% was achieved for the generated radiation.
\end{abstract}
\end{frontmatter}

\section{Introduction}
Microlens arrays have found host of applications in linear 
optics, mainly related to the wavefront profile reconstruction problem
\cite{Hartman}.
However, the potential in nonlinear optics of microlens 
arrays has not been fully exploited yet. To date, microlens arrays were 
employed to obtain three-dimensional images of biological samples in a 
multiphoton excitation microscope\cite{Bewer98,Fujita00}. 
More recently, a nonlinear all-optical switch based on the Talbot effect and 
the spatial soliton\cite{solitons} excitation process was implemented with a lenslet array
\cite{Talbot}. 

Here we show how a microlens array can be employed successfully in combination with a quadratic nonlinear crystal to get a 
compact, high-quality, ultrashort, parametric light source. 
In the design of pulsed optical parametric generators (OPG), one of the 
problems to solve is that of achieving simultaneously good energy stability 
and high beam-quality. Actually, while the first feature is 
achieved in the regime of gain saturation, the latter is obtained in the 
conditions of linear amplification.
To avoid the problem, the usual approach is that of implementing 
several stages of spatial filtering and amplification in the OPG, which makes the device rather complex.
A possible way to overcome these limitations and improve spatial quality 
while preserving the energy stability in OPGs is to exploit the features of 
the parametric spatial soliton \cite{DiT98a}, for which the beam quality is ensured thanks to the occurence of a suitbale spatial mode-loking effect \cite{DiT98a,DiT98c}.
In fact, spatial solitons exhibit excellent beam quality (diffraction limit \cite{DiT98a}) in all the trapped wavelengths while their energy content is directly proportional to that of the input pump\cite{Torner99}. 
However, due to the finite damage threshold of nonlinear materials, the 
energy that can be converted into solitons is quite small, of the order of 
few hundreds of nJ/ps.
To increase the overall energy conversion, a multi-pump (multi-pixel) scheme 
has been tested \cite{Min00} in conditions where the single pump 
beam is able to excite a spatial soliton.
The conversion efficiency of the device was however limited to a few percent, mainly because of the losses caused by amplitude-modulation mask that we used for the multi-pump beam generation. 

In the present work, we show that a dramatic improvement of the conversion
efficiency in the multi-pixel OPG (MPOPG) is possible by using a phase mask
(\textit{i.e.} a microlens array), instead of an amplitude one, to multiplex the
input pump beam.
We show that our MPOPG is able to achieve high level and stable
energy conversion while preserving locally high spatial quality all over the 
tuning range. We mention that the tunability and the spectral features of the parametrically generated spatial solitons, of paramount importance for the application, have never been presented before. 

The paper is structured as follows. In the first section, we describe the experimental set-up and report on the spatial features of the output radiation, providing evidence of the formation of parametric spatial solitons. 
In the next section, we discuss how the tuning conditions affect the soliton pattern and explain why the localization of the pump beam seems to disappear, while the signal field retains the soliton-like features. Another section is devoted to the features of the entire source as a whole. Issues like the net conversion efficiency, energy fluctuation and the spectral features are addressed. Finally, conclusions are drawn together with a proposal for a possible exploitation of the multipixel source as a practical device. 

\section{Spatial features of an array of solitons}

\begin{figure}
\epsfig{file=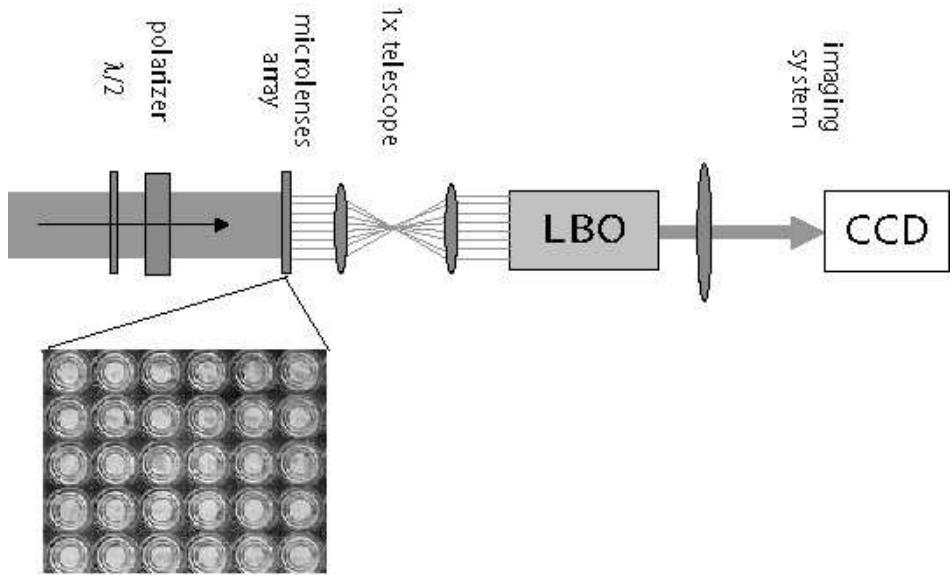, width=12.5cm}
\caption{\label{setup}Sketch of the experimental set-up. In the inlet, a picture of the microlenses array.
The second harmonic of our amplified laser source is regulated by means of a $\lambda/2$ plate combined with a linear polarizer.}
\end{figure}

The experimental setup, depicted in Fig. \ref{setup}, is based on a microlens array that devides an intense high-frequency pump beam into a matrix of focused beams impinging on a quadratic nonlinear crystal. The pump is a collimated gaussian beam of 3 mm $1/e^2$ radius 
from a frequency doubled Nd:glass pulsed laser source (TWINKLE, Light Conversion Ldt., delivering 1 ps pulses at the wavelength of 527.5 nm). The central part of the beam was selected by means of a 2.5 mm round aperture placed short before the microlens array.  
Such cut beam contains about 20\% of the total energy and provides a quasi flattop beam profile.
The fused silica microlens array (WaveFrontSciences Inc., see inlet in Fig.1) is composed by a square-lattice matrix of 83x102 microlenses. The focal length of each lens is 4.6 mm. The pitch of the array is of 108 $\mu$m while the measured spot diameter in the focal plane is of about 15$\mu$m FWHM. 

The focal plane of the microlens array is imaged on the
input face of a 15-mm-long lithium triborate crystal (LBO) by means of a 2-lens
telescope of magnification 1. This has been necessary since the oven containing the crystal has rather long edges that forbid to mount the array directly in front of the crystal. 
A single-lens imaging-system placed behind the nonlinear crystal is used to reconstruct on a CCD camera the beam profile at the exit face of the crystal.

\begin{figure}
\mbox{\epsfig{file=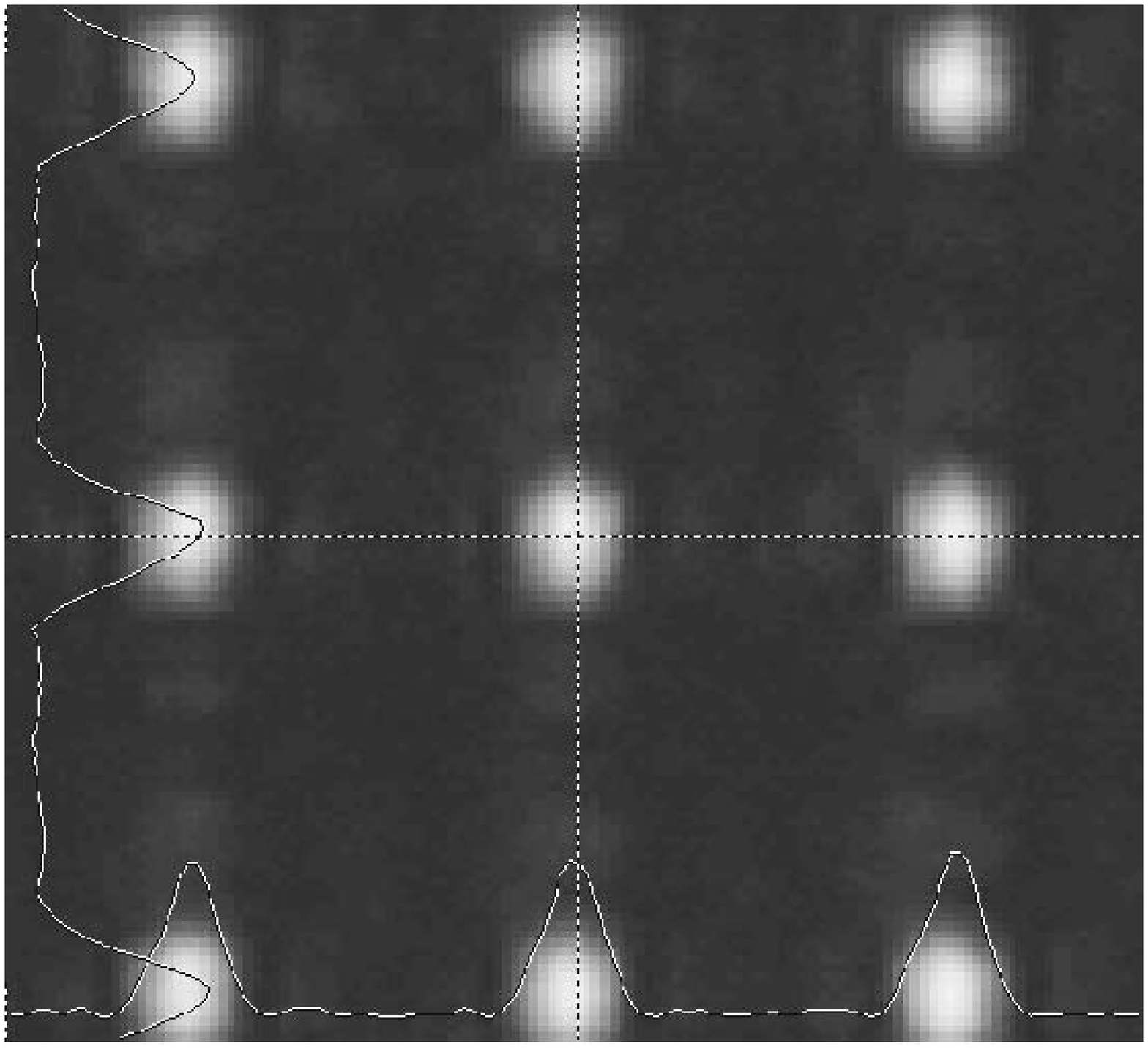, width=3cm}\quad
\epsfig{file=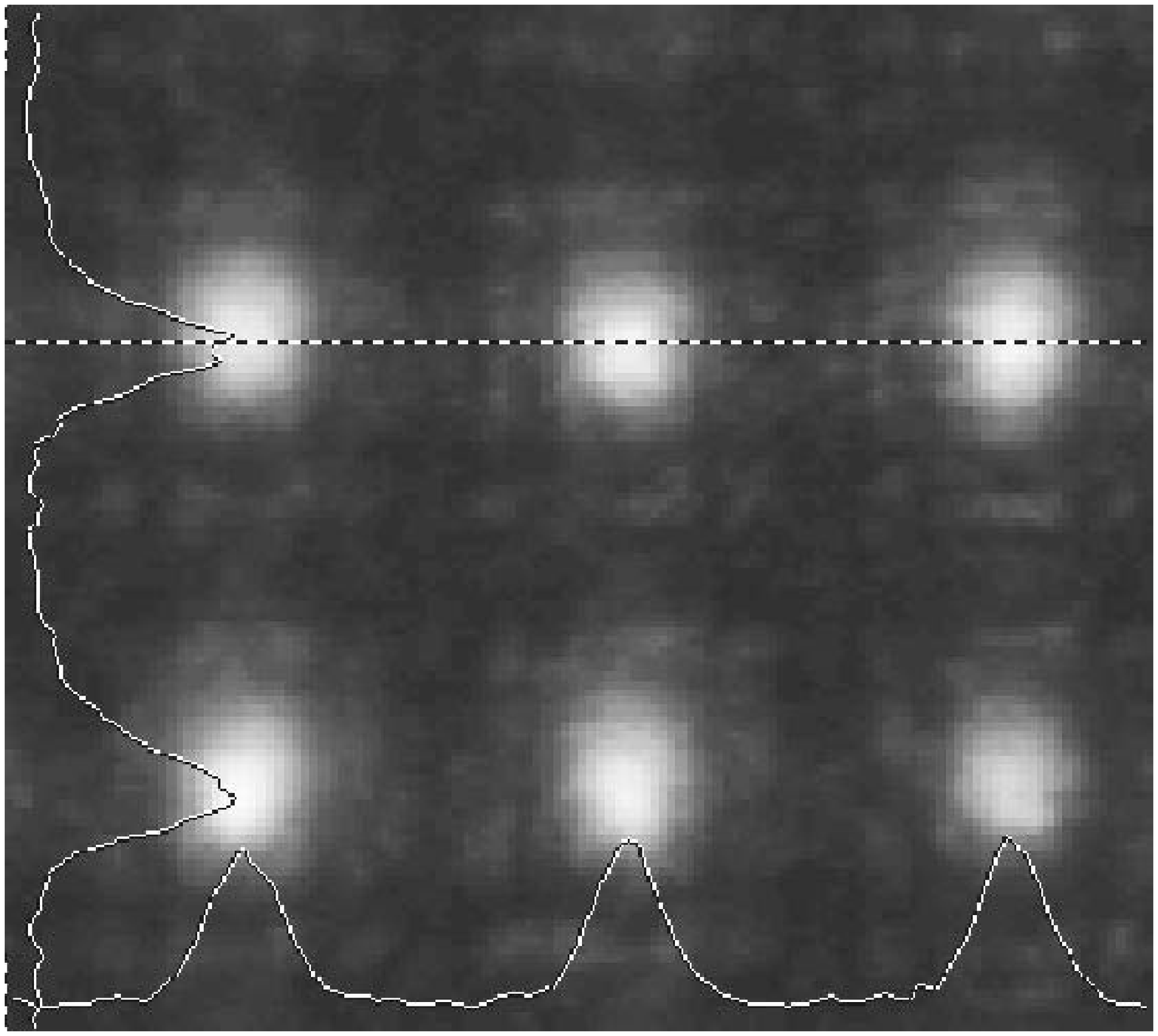, width=3cm}\quad
\epsfig{file=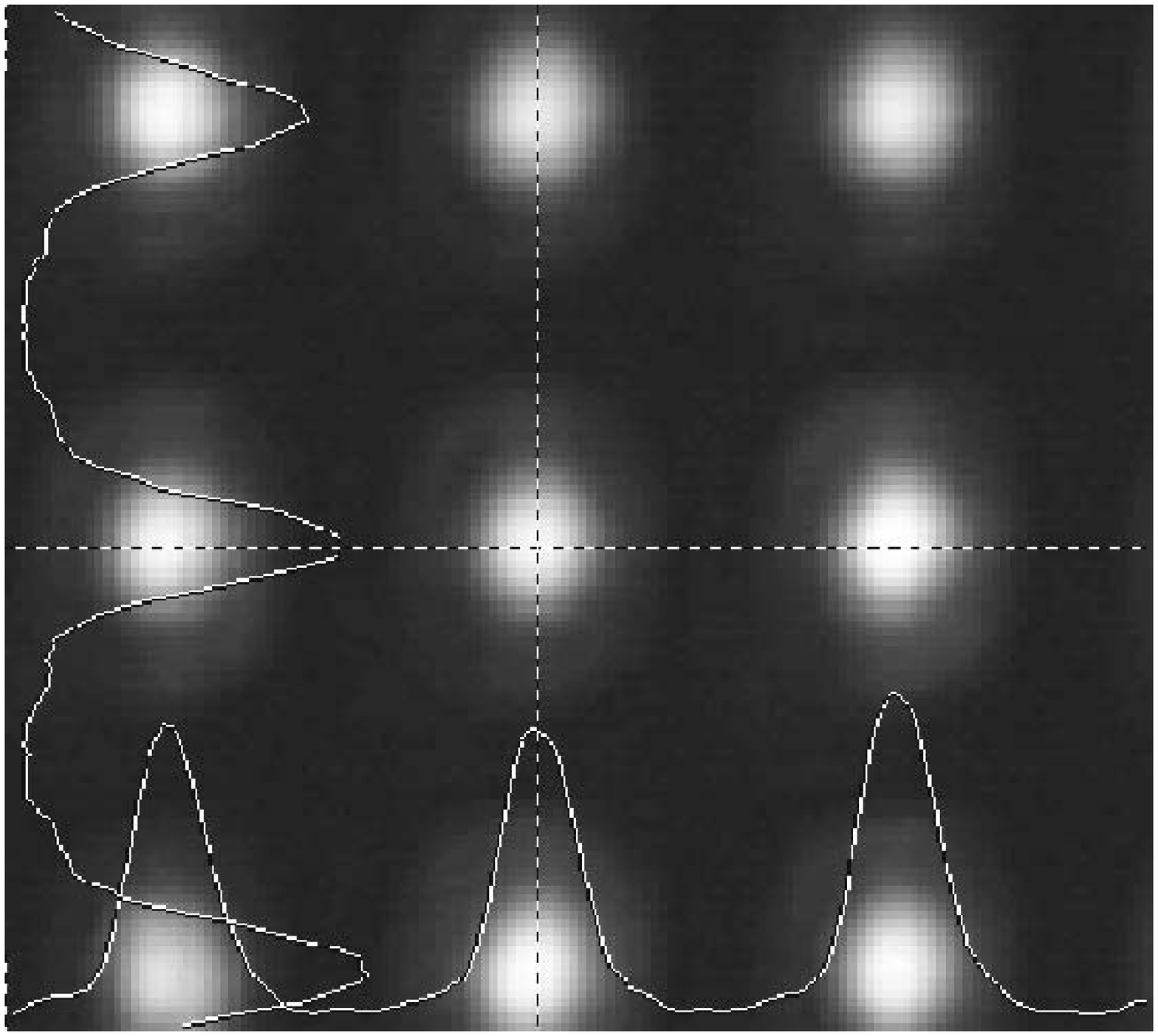, width=3cm}\quad
\epsfig{file=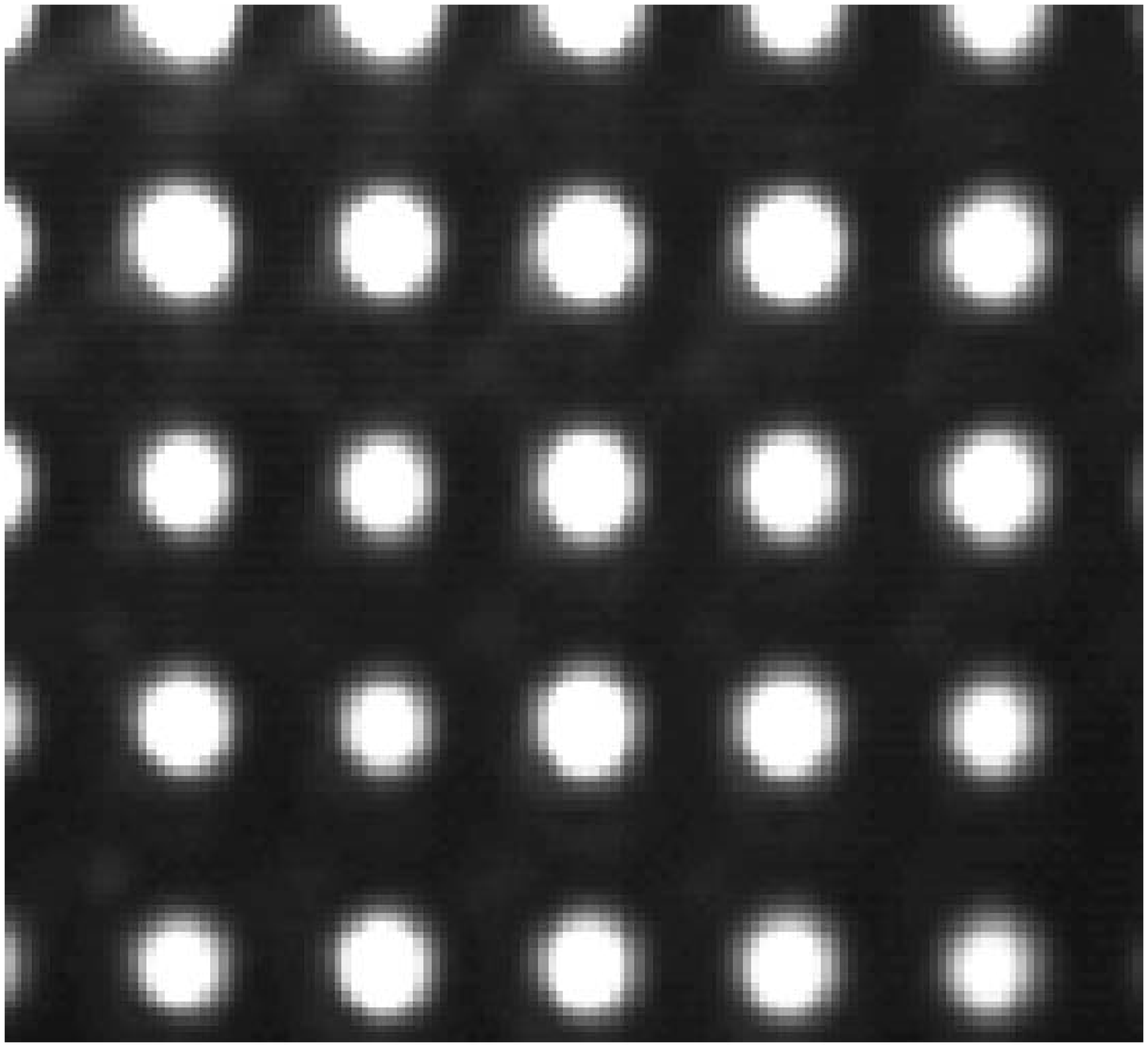, width=3cm}}
\caption{\label{nearfield} Fluence profiles of a small area of: a) The exciting pump array at the LBO crystal entrance ($\lambda=527.5$ nm); b) the trapped pump at the output face (T=122$^\circ$C, $\lambda_s=790$ nm); and c) the signal wave profile at the degenerate phase matching condition (T=164$^\circ$ C, $\lambda_s=1055$ nm). Transverse profile sections referring to the cursor position are plotted on the side of each figure, for a more quantitave estimation of the actual beam shape. The center of the cursor in figures a), b) and  c) correspond to the same optical axis. In Fig. d) a saturated image of the signal (conditions as in c)) to show that the background level of the pattern is extremely low.}
\end{figure}

In the experiments, the crystal temperature was tuned from 108$^\circ$C (a condition that has already been proven to be the best for the excitation of parametric spatial solitons from quantum noise fluctuations \cite{DiT98b}) up to 164$^\circ$C, which corresponds to the degenerate case. In this interval, the generated signal central wavelength ranged continuously from $\lambda_s=730$ nm to $\lambda_s=1055$ nm.
In the temperture region close to the edge of the tuning (from 108$^\circ$C to 125$^\circ$C), narrow, mutually-trapped beams of high spatial-quality appear in both the pump and signal fields (Fig. \ref{nearfield}.b, for the pump beam, see also \cite{Min00}), as the input pump intensity is raised above the level of 0.8 GW/cm$^2$ (measured before the microlens array). The same high quality of the signal is maintained up to the degeneracy point (see Fig. \ref{nearfield}.c).The position of each beam corresponds to that of the input pump pixels (Fig. \ref{nearfield}.a), projected on the output face along the propagation direction of the pump radiation, this being a typical signature of the achievement of the spatial soliton regime \cite{Min00,Bra01}. 
As expected, the contrast of the pattern is much higher in the signal than in the pump field, an effect related to the exponential trend of the parametric amplification. Saturared images of the signal beam show that the background level is in fact negligible (Fig. \ref{nearfield}.d). 

\begin{figure}
\epsfig{file=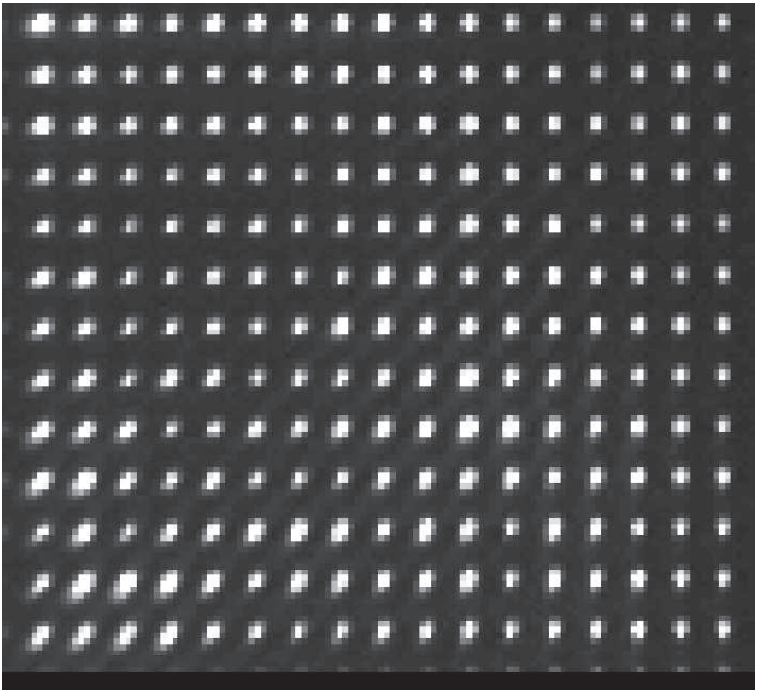, width=10cm}
\caption{\label{largearray} A large area shot of the output signal radiation. Inhomogeneities are evident, mainly connected with defects on microlenses array and crystal surfaces. Experimental conditions: crystal temperature = 164$^\circ$C, pump intensity = 0.8 GW/cm$^2$ (before lenslet array), emitted signal wavelength = 1055 nm}
\end{figure}

The main advantage of using a microlens array is that they allows for an efficient multiplexing of the pump beam, therefore arrays made-up of several hundreds of spatial solitons could be easily excited with a single laser pulse (see Fig. \ref{largearray}).
Each soliton of the array has been already proved to be independent from its neighbours \cite{Min00,Bra01}, therefore arbitary patterns of solitons might be easily excited by allowing only selected microlenses to shine on the nonlinear crystal.

The overall conversion efficiency of the system (energy of the parametric radiation/input pump energy) has been greatly improved respect to the scheme employing an amplitude-transmission mask \cite{Min00}. Conversion efficiencies up to 32\% have been measured.
This is actually a high figure, if one considers that it was achieved with a single-pass scheme. Notice that our scheme, besides being extremely compact, does not introduce any relevant loss into the system. More details on the conversion efficiency issue are given in section \ref{conversion}.

As for the uniformity of the array pattern, Fig. \ref{largearray} shows that it is quite satisfactory. 
The statistic carried out over 221 pixels shows that the rms fluctuation of 
the peak intensity over the frame is $\pm$ 24\% around the mean value. The observed inhomogeneities could be easily related to  defects on the lenslet array, defects of the crystal and to spatial inhomogeneity of the pump beam before the lenslet array.
The last one is the most detrimental for the homogeneity, if no flat-top beams were available to shine on the microlens array. However, special schemes as that described in \cite{arrays} might be exploited to enhance the pixel-to-pixel uniformity of the input pump.   

For what concerns the single pixel energy content, from the peak intensity
of the incoming pump it can be estimated that each lenslet collects at its 
maximum about 0.6$\mu$J in case of pump well within the saturation range. 
Allowing a 30\% conversion efficiency, each pixel carries about 0.2$\mu$J of 
signal and idler radiation.

\section{Parametric spatial solitons and wavelength tuning}

An essential feature of a parametric light source is the tunability, therefore the further step of our investigation has been a careful characterization of the parametric emission over the same tuning range of the LBO crystal. 

As mentioned, well contrasted arrays of beams like in Fig. \ref{nearfield}.c were observed to be the typical output of the source in all the tuning range, although the single beam diameter was observed to increase as the degeneracy condition was approached (from 20 to 30 $\mu$m FWHM). 
Interestingly, this feature is not apparently preserved for the high-frequency pump field. In fact, while for temperatures below 125$^{\circ}$C (corresponding to $\lambda_s=805$ nm) a nice array of self-trapped beams is still observed, no clear signature of spatial trapping of the pump field could be detected in conditions closer to degeneracy.

This puzzling observation has been recently explained in the frame of a space-time model of the parametric interaction \cite{Paolo03}. With pulsed light, only the central part of the exciting pulse can form a self-trapped, almost non-diffracting beam \cite{Simos02}. The tails of the pulse are too weak to initiate the self-focusing process and therefore contribute to the linerly diffracting background.
However, the mean intensity of the self trapped part of the pump pulse becomes smaller and smaller as the degeneracy tuning condition is approached. Therefore, on time integrated pictures of the pulse, the self-trapped component of the pump pulse seems to vanish into the linearly diffracting background. On the contrary, the intensity of the self-trapped signal (idler) beam increases, so that apparently the soliton survives only in the signal and idler fields. 
The reason of this behaviour is linked to the parametric amplification bandwidth of the nonlinear crystal and on the dynamics of the amplification of noisy fields. This carries some analogy with recent studies on the 3D space-time modulational instability and focus wave modes generation in parametric systems \cite{Trillo02,Orlovas02}.

Let us point out that this effect has however no relevant impact on the operation of the soliton array as a parametric light source. In fact, the most important feature is that the signal field output is still structured as a regular and well contrasted array of beams of high spatial quality.
Moreover, the observed color unbalance actually contributes to achieve high conversion efficiencies figures. 

\section{Global features of the source}
\label{conversion}

Having described the local features of the parametric light source, we turn now our attention to the global features of the generator, such as the total conversion efficiency, the average energy fluctuation, and the spectral features.

\begin{figure}
\epsfig{file=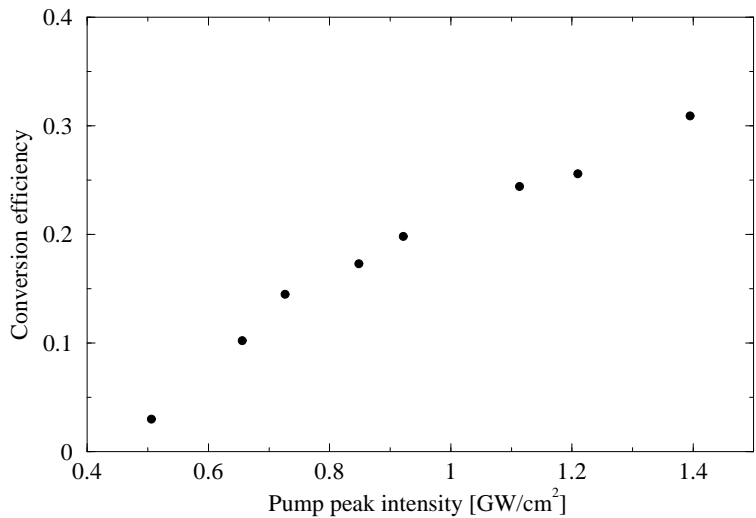, width=10cm}
\caption{\label{Fconversion}Overall conversion efficieny at degeneracy as a function of the input pump intensity.}
\end{figure}

\begin{figure}
\epsfig{file=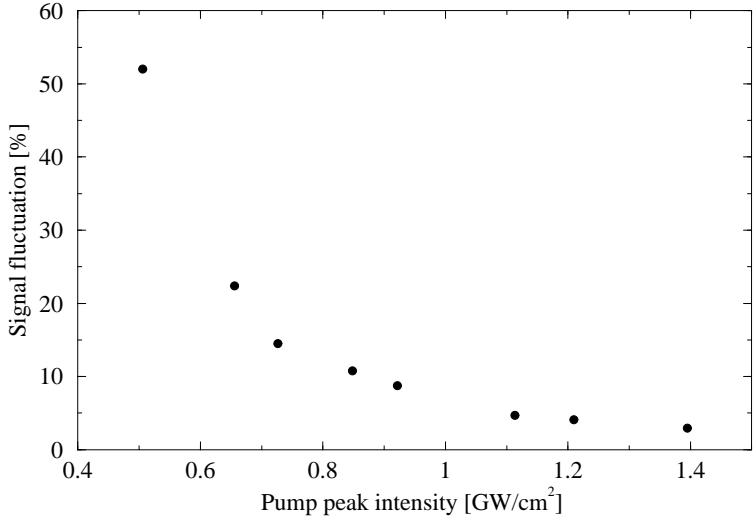, width=10cm}
\caption{\label{fluctuate}Rms fluctuation of the emitted parametric radiation as a function of the input pump intensity, for the same operating conditions as in Fig. \ref{Fconversion} (see also text).}
\end{figure}

In figures \ref{Fconversion} and \ref{fluctuate} the conversion efficiency and the energy rms fluctuation of the overall generated radiation at degeneracy (T = 164$^\circ$C) are plotted as a function of the input peak intensity.
These parameters are to be considered an average value over the whole
output, since they were measured by simply collecting the output 
down-converted radiation (the pump was rejected by means of a dielectric 
mirror) on a large area pyroelectric energy meter. 
The graph shows that raising the pump intensity above 0.8 GW/cm$^2$ (measured before the microlens array), the conversion efficiency shows a tendency to saturate towards a limit of 35\%. Notice that we could not fully enter the saturation regime with the available power. The maximum achieved conversion efficiency is slightly above 30\%. In the same saturation regime, the shot-to-shot energy fluctuations were observed to drop down to an asimptotic limit of about 3\%, which is less than two times that of the input pump energy (measured to be 1.7\%, as illustrated in Figure \ref{fluctuate}). 
Similar trends and figures for the energy conversion and the energy fluctation were observed over the whole tuning range of the crystal, therefore confirming our scheme as a promising prototype of a novel parametric light source.

\begin{figure}
\epsfig{file=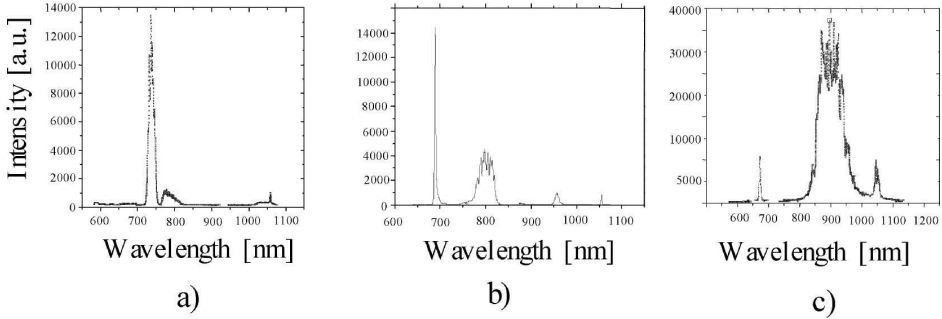, angle=90, width=12.5cm}
\caption{\label{spectra} Spectra of the signal radiation for different tuning conditions of the LBO crystal. a) T=108$^\circ$C; b) T=129$^\circ$C; c) T=159$^\circ$C. The enlargement of the spectral bandwidth is evident, as the degeneracy condition is approached.}
\end{figure}

A spectrograph was employed to analyze the output signal radiation spectra.
Fig. \ref{spectra} shows a selection of spectra averaged over 15 shots and taken at different tuning conditions. As expected from the dispersive properties of our type I LBO crystal, the spectra get very broad on approaching degeneracy, their FWHM bandwitdh ranging from 20 nm, at the very edge of the tuning (see Fig. \ref{spectra}.a, left spike) to about one hundred nm close to degeneracy (Fig. \ref{spectra}.c). Single-shot measurements reveal that the spectra are deeply modulated, their actual pattern changing form laser-shot to laser-shot.  
These data suggest that the coherence time of the output parametric radiation could be quite low. Consistently, space-time simulation of the system revealed that the signal (pump, idler) pulse should exhibit a rugged temporal profile, with details in the few fs scale \cite{Paolo03}.
This low temporal coherence, that might be considered detrimental in some application, can turn into an advantage for such applications where intense, incoherent light is needed (see also next section).
We point out that extremely high intensities could be achieved instantaneously inside the signal pulse (in the order of several hundreds of GW/cm$^2$), a non common feature among the incoherent light sources. Moreover, the excellent spatial quality of the signal beams let us forsee a high focusability of such beams, that might provide even higher peak intensities.

In case narrow-band, long coherence-time parametric radiation is needed, the scheme has to be implemented with a narrow bandwidth seed injection at the signal (idler) wavelength. Measurements (data not shown) prove that close to degeneracy the bandwidth of the signal can be as narrow as 4-6 nm, in case a weak, seed is injected (of energy about 1000 times smaller than that of the pump) in the crystal with the intense pump beam.
 
\section{Conclusions and perspectives}

In conclusion, we have shown that a microlens array can be succesfully employed to multiplex an intense pump beam and excite a large array of parametric spatial solitons in a quadratic nonlinear material. We investigated the spatial and the temporal properties of the emitted signal radiation and showed that the radiation can be tuned easily without affecting significantly its spatial quality.

From the data reported above, it is evident that the multi-beam pumping
couples stability and high conversion efficiency figures with high spatial 
quality of the output radiation.
These characteristics are not easily obtained in conventional OPGs, which need 
several passes and spatial filtering to obtain high spatial-quality output. 
The main advantage of the proposed scheme is its intrinsic compactness, which 
opens new persectives in the development of OPGs whose dimensions are in the 
order of few cubic centimeters.

For what concerns the spectral quality of the output radiation, the 
amplification bandwidth of the parametric process in the single pass scheme
is quite large (several tens of nm), therefore the output radiation is expected to feature coherence times in the few fs range.   
This quality makes the MPOPG a unique source of intense temporally (quasi) incoherent but spatially highly coherent, with an average power-per-pixel (about MW) orders of magnitude higher than any currently available broadband tabletop source \cite{superLED,holyfiber}. Notice that the high spatial quality of each beamlet composing the array can be exploited to obtain even more focused beams and therefore extremely high peak intensities (tens of TW/cm$^2$).

On the contrary, higly temporal coherent radiation could be easily produced by the source by merely adding a single seeding pass to the amplification scheme\cite{DiT98b}. 
Moreover, the intrinsic multicannel structure of the MPOPG could be exploited to build-up an extremely flexible all-optically reconfigurable light source. For instance, patterns of parametric solitons might be excited by letting the pump mix with a spatially modulated seeding beam \cite{CLEO99,Talbot}, while a frequency dispersed seeding might be useful to excite an array in which each pixel has a different central wavelength. This last scheme could be useful, for instance, in spectroscopic measurements (linear and nonlinear) of spatially extended or patterned chemical/biological samples. 

All authors aknowledge Instituto Nazionale per la Fisica della Materia, MIUR (COFIN01 and FIRB01), the Department of Sciences of the University of Insubria (``Young reserchers project'') and the European Commission (CEBIOLA contract ICA1-CT-2000-70027) for financial support.
All the authors are thankful to G. Arrighi and G. Blasi for helping in the 
arrangement of the experiments.

\end{document}